\documentclass[letters]{aa} 

\usepackage{epsfig}     
\usepackage{graphicx,color}     
\usepackage{amssymb}            
\usepackage{url}                
\usepackage{amsmath}            
\usepackage{rotating}                   
\usepackage{float}                      
\usepackage{textcomp}           
\usepackage{epstopdf}
\usepackage{dcolumn}
\usepackage{times}
\usepackage{tabularx}
\usepackage{hyperref}
\hypersetup{
    colorlinks,
    citecolor=blue,
    filecolor=blue,
    linkcolor=blue,
    urlcolor=blue,
    menucolor=black}
\usepackage{ulem} 
\usepackage[english]{babel}
\usepackage{booktabs}
\usepackage{gensymb}
\usepackage{subfig}
\usepackage{appendix}

\usepackage{etoolbox}

\makeatletter
\renewcommand*\aa@pageof{, page \thepage{} of \pageref*{LastPage}}
\makeatother

\begin{document}
\title{Data-constrained 3D MHD Simulation of a Spiral Jet Caused by an Unstable Flux Rope Embedded in Fan-spine Configuration}

\author{Z. F. Li\inst{1,2,3}, J. H. Guo\inst{1,5}, X. Cheng\inst{1,2,3}, M. D. Ding\inst{1,2}, L. P. Chitta\inst{3}, H. Peter\inst{3,4}, S. Poedts\inst{5,6}, D. Calchetti\inst{3}}

\institute{School of Astronomy and Space Science, Nanjing University, Nanjing, 210046, People's Republic of China\\
\email{xincheng@nju.edu.cn}
\and Key Laboratory of Modern Astronomy and Astrophysics (Nanjing University), Ministry of Education, Nanjing 210093, China
\and Max Planck Institute for Solar System Research, Justus-von-Liebig-Weg 3, 37077 G{\"o}ttingen, Germany
\and Institut f{\"u}r Sonnenphysik (KIS), Georges-K{\"o}hler-Allee 401A, 79110 Freiburg, Germany
\and Centre for Mathematical Plasma Astrophysics, Department of Mathematics, KU Leuven, Celestijnenlaan 200B, B-3001 Leuven, Belgium
\and Institute of Physics, University of Maria Curie-Skłodowska, Pl.\ Marii Curie-Skłodowskiej 5, 20-031 Lublin, Poland}

\date{received ...; accepted ...}

\abstract{Spiral jets are impulsive plasma ejections that typically show an apparent rotation motion. Their generation, however, is still not understood thoroughly. Based on a high-resolution vector magnetogram from the Polarimetric and Helioseismic Imager onboard Solar Orbiter, we construct a data-constrained three-dimensional (3D) MHD model, aiming to disclose the eruption mechanism of a tiny spiral jet at a moss region observed on March 3 2022. The initial configuration of the simulation consists of an extrapolated coronal magnetic field based on the vector magnetogram and an inserted unstable flux rope constructed by the Regularized Biot–Savart Laws method. 
Our results highlight the critical role of the fan-spine configuration in forming the spiral jet and confirm the collapse of the pre-existing magnetic null to a curved 3D current sheet where external reconnection takes places. It is further disclosed that the flux rope quickly moves upward, reconnecting with the field lines near the outer spine, thereby enabling the transfer of twist and cool material from the flux rope to the open field, giving rise to the tiny spiral jet we observed. The notable similarities between these characteristics and those for larger-scale jets suggest that spiral jets, regardless of their scale, essentially share the same eruption mechanism.}

\authorrunning{Li et al.}
\titlerunning{Data-constrained 3D MHD Simulation of a Spiral Jet}

\maketitle

\section{Introduction}
Coronal jets are transient collimated outflows that appear bright in the extreme ultraviolet \cite[EUV;][]{Wang1998} and soft X-ray \cite[SXR;][]{Shibata1992} observations. They are of significant interest as they can transfer mass and energy to the outer corona and solar wind (\citealt{Priest2021} and \citealt{Shen2021}). With recent high-resolution observations, coronal jets are even found to share a similar physical origin with large-scale solar eruptions like coronal mass ejections \citep{Sterling2015, Liujj2015, Joshi2020b} and may have a contribution to partial acceleration in solar wind \citep{Moore2011, Chitta2023}.

Magnetic reconnection, a fundamental process in magnetized plasma that converts magnetic energy to other forms, is assumed to be the key driver of coronal jets. It has been widely studied in observations and theoretical models \citep{Shibata1992, Shibata2007, Schmieder2014, Wyper2016, Tian2018}. Variations in the morphology of jets as observed in SXR and EUV images, along with the inferred sudden changes in magnetic connectivities apparent from the changes in emission patterns, have been extensively used as evidence for reconnection. \cite{Innes1997} reported spectral evidence that shows red and blue shifts in Si IV 1393~\AA\ line profiles, indicating the presence of bidirectional jets ejected from reconnection sites. Additionally, \cite{Shimojo2000} reported an SXR jet with a morphology closely related to micro-flares generated by magnetic reconnection. 

Inferring magnetic reconnection based solely on observations alone is not straightforward. MHD simulations provide a robust approach to exploring the reconnection and the relation to coronal jets. \cite{Yokoyama1996} conducted a 2D MHD simulation, demonstrating that jets can be produced when newly emerging magnetic flux reconnects with the oblique background field. Subsequent studies found that 3D reconnection between the twisted core field and ambient open field can generate spiral jets in coronal holes \citep{Pariat2009, Pariat2010}. Recently, \cite{Wyper2019} simulated a spiral jet and found that it is caused by a combination of magnetic breakout and kink instability of the pre-existing flux rope, which is formed through imposing artificial surface motions at the bottom boundary. 

Complementary to MHD simulations, observational data-constrained or data-driven simulations have become a promising method for reproducing coronal jets and even more complex events. \cite{Cheung2015} presented time-dependent magneto-frictional data-driven simulations of recurrent jets originating from a fan–spine magnetic system. \cite{Nayak2019} demonstrated the mini-filament breakout model using a 3D data-constrained MHD simulation. This approach also facilitates a more detailed analysis of thermodynamic and topological properties \citep{Guo2023}. Additionally, \cite{Jiang2016} identified the positive feedback between emerging flux and external magnetic reconnection as a crucial factor for the evolution transiting to the eruptive state.

Recently, \cite{Cheng2023} reported a new observation of persistent null-point reconnection in the EUV at the smallest spatial scale (about 390~km) to date, during which an interesting spiral jet was also observed, lasting for 10 minutes and being identified to arise from a mini-filament eruption. Based on EUV images from the High-Resolution Imager (HRI$_{EUV}$) at 174\,\AA\ of the Extreme-Ultraviolet Imager \citep[EUI;][]{Rochus2020} on board the Solar Orbiter \citep{Muller2020}, the jet of interest, originating from the tiny null point configuration, was found to closely resemble spiral jets typically observed \citep{Raouafi2016}, albeit on a smaller scale. To further understand the generation of this tiny spiral jet and unresolved transferring of the mass and twist from the lower atmosphere to the corona in observations, we employ observational data-constrained simulation using the Message Passing Interface Adaptive Mesh Refinement Versatile Advection Code \cite[MPI-AMRVAC;][]{Keppens2003, Keppens2012, Xia2018}, the initial condition of which is well constrained by the high-resolution vector magnetogram up to now provided by the Polarimetric and Helioseismic Imager \cite[SO/PHI;][]{Solanki2020} onboard Solar Orbiter. 
Data and numerical methodology are described in Section~\ref{data}. Numerical results are presented in Section~\ref{results}, followed by a summary and discussions in Section~\ref{summary}.

\begin{figure*}[!ht]
    \centering
    \includegraphics[width=16cm]{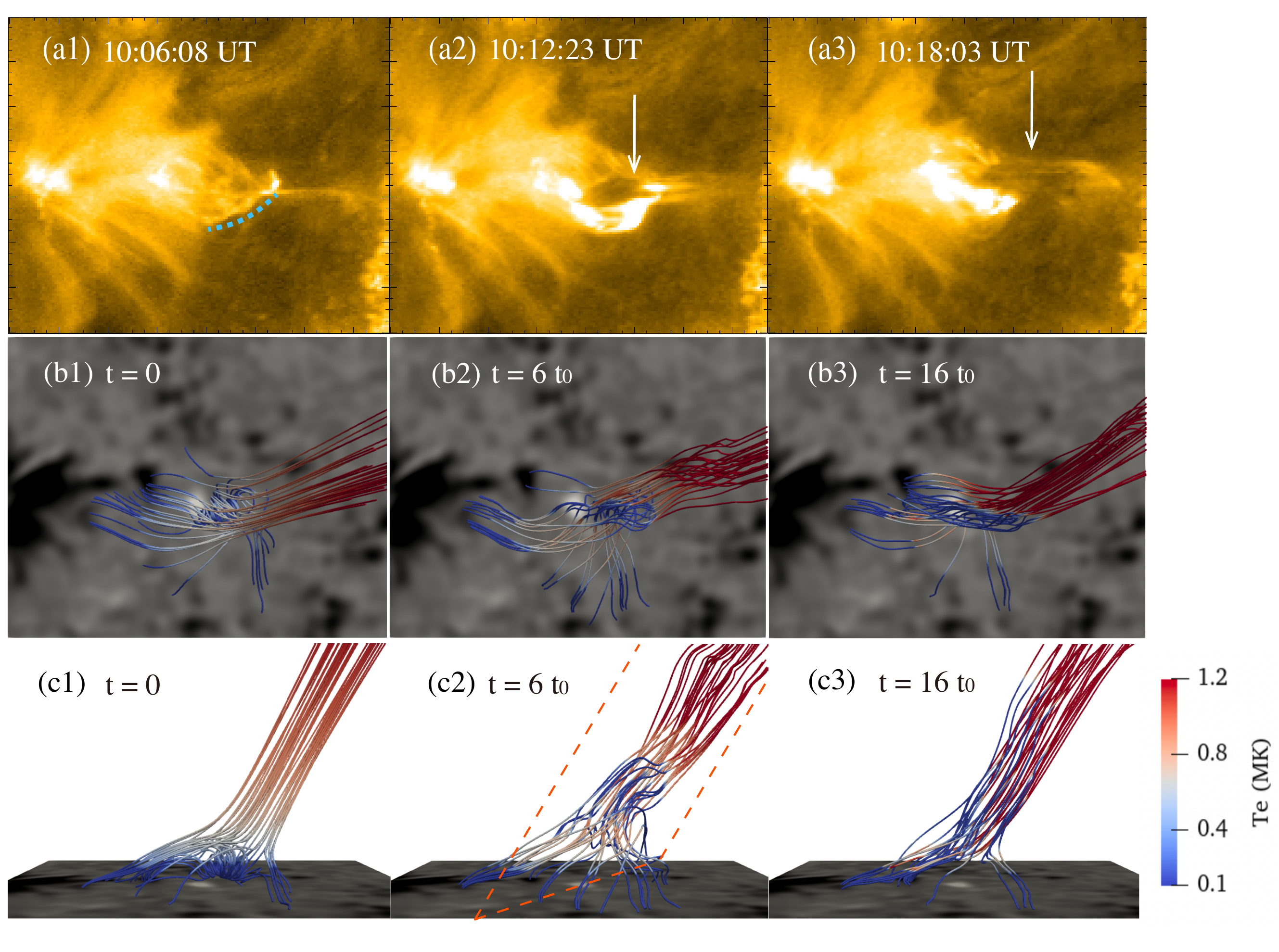}
    \caption{Comparison between observations and simulations at three instances. (a) EUI 174 {\AA} images displaying the temporal evolution of the spiral jet eruption with arrows indicating dark threads. The blue dashed curve in panel (a1) indicates the path of the inserted flux rope. (b) Temporal evolution of 3D magnetic field lines as viewed from the top. (c) The same as panel (b) but from the side view. The lines in blue and red correspond to cold and warm plasma, respectively. The red dashed lines indicate the plane in Figure \ref{figa2}.} 
    \label{fig1}
\end{figure*}

\begin{figure*}[!ht]
    \centering
    \includegraphics[width=14cm]{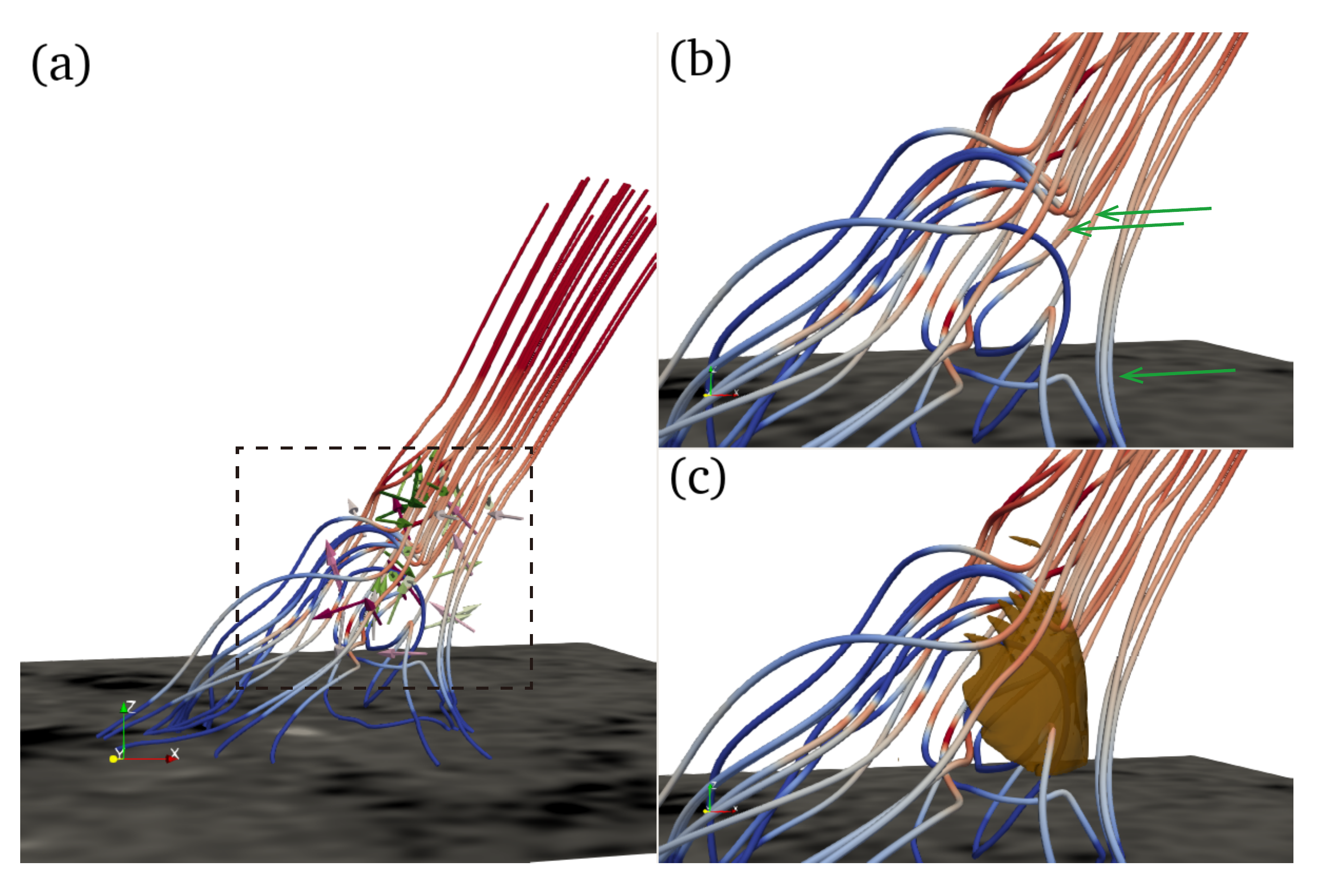}
    \caption{3D configuration of external reconnection in our simulation at $t = 3 t_0$. (a) Overview of magnetic field lines with the velocity vectors denoted by the arrows. The colors green and pink represent the positive and negative of $v_x$, respectively. The colors of field lines have the same meaning as Figure \ref{fig1}. (b) Zoomed-in reconnection region. The green arrows indicate magnetic dips. These heated, curved field lines are formed through interchange reconnection between the erupting flux rope and the overlying open fields. (c) The same region as in panel (b) but with the 3D curved current sheet.}
    \label{fig2}
\end{figure*}

\section{Observation and Numerical Setup}
\label{data}
\subsection{Instruments}
We use the calibrated L2 174~\AA\ images on March 3 2022, obtained from HRI$_{EUV}$ onboard Solar Orbiter. At that time, Solar Orbiter was located at a distance of 0.50~AU from the Sun and almost aligned with the Sun-Earth line, with an angular separation of approximately 7 degrees. The dataset \footnote{https://doi.org/10.24414/2qfw-tr95} was taken between 09:40 UT and 10:40 UT with a cadence of 5\;s. The pixel size is 0.492$''$, corresponding to a spatial scale of about 195 km.
The vector magnetogram used in the magnetic field extrapolation and numerical simulation was acquired by the High-Resolution Telescope \cite[SO/PHI-HRT;][]{Gandorfer2018, Solanki2020} of PHI onboard Solar Orbiter. The dataset spans the same observational period as EUI, with a temporal cadence of 300 s. The images have a spatial resolution of 0.5$''$ per pixel, corresponding to $\simeq 200\;$km on the Sun. They are calibrated using the SO/PHI-HRT on-ground pipeline \citep{Sinjan2022,Calchetti2023}. The ambiguity in the transverse component of the magnetic field is removed using an adapted version of the ME0 method \citep{Metcalf1994} as implemented in the pipeline of the Helioseismic and Magnetic Imager \cite[HMI;][]{Scherrer2012} onboard Solar Dynamics Observatory \citep[SDO;][]{Pesnell2012} \citep{Metcalf2006,Hoeksema2014}.

We remove the jitter in both datasets using a cross-correlation method (see Appendix A.1 in \cite{Chitta2022} for details). Note that these two datasets' fields of view (FOVs) are not identical. We adopt a cross-correlation approach to coregister the images of the HRI$_{EUV}$ and HRT. Following the methodology outlined by \cite{Kahil2022}, we initially aligned the HRI$_{Lya}$ image with the HRI$_{EUV}$ image and then aligned the HRT image with the HRI$_{Lya}$ image. HRI$_{Lya}$ images the Sun in Ly-a at 1215~\AA\ and shows the chromosphere. This procedure ensures that the datasets of the HRT and HRI$_{EUV}$ are accurately co-aligned. Upon examining the coaligned images, we find a good match in position between the brightenings and regions with strong magnetic fields, as evident from the comparison between Figure~1 in \cite{Cheng2023} and Figure~\ref{figa1}(a).

\subsection{Overview of Observations}

Observations from multiple EUV passbands of the Atmospheric Imaging Assembly \citep[AIA;][]{Lemen2012} onboard SDO have been extensively analyzed by \cite{Cheng2023}. Here, we give an overview of this jet event as observed by Solar Orbiter. This tiny spiral jet, arising from a moss region between two active regions, began around 10:11 UT on March 3, 2022, and lasted for approximately 10 minutes, featuring a spiral motion of dark and bright threads. The zoomed-in LOS magnetogram reveals that the jet is located above a small positive polarity region embedded within a predominantly negative polarity region (Figure~\ref{figa1}(b)). This configuration suggests a fan-spine configuration, where the spine connects the inner positive polarity enclosed by the fan, with footpoints rooted in surrounding negative polarities as confirmed in \cite{Cheng2023}. This configuration remained relatively unchanged throughout the process despite some disturbances.
Figure \ref{fig1}(a1)-\ref{fig1}(a3) shows the evolution of the spiral jet. As suggested by \cite{Cheng2023}, the magnetic reconnection continuously occurred at the null point before the jet, rendering a faint jet spire visible before the eruption. During the eruption, the mini-filament (threaded by a sheared magnetic field, e.g. \cite{Sterling2015}) began to ascend rapidly, resulting in a spiral jet characterized by fragmented cool plasma ejecting and propagating along the outer spire towards the farther end.

\subsection{Adiabatic MHD model}
\label{simulation}

Based on the SO/PHI vector magnetogram, we perform a nonlinear force-free field (NLFFF) extrapolation to investigate the magnetic topology structure of this jet, serving as the initial magnetic fields of the following MHD simulation. To account for the presence of the mini-filament indicated by observational features, in the initial field, we manually insert a flux rope,  which cannot be captured by the NLFFF extrapolation. 
The twisted flux rope is reconstructed by the Regularized Biot–Savart Laws (RBSL) method  \citep{Titov2018}. Four key parameters govern this method: the flux rope path, flux rope radius, $a$, electric current $I$, and toroidal magnetic flux, $F$. We outline the projected flux rope path by following the mini-filament spine as indicated by the blue dashed curve in Figure~\ref{fig1}(a1). This path is selected based on observations of the dark lane in this region, and the resulting extrapolations yield a dome structure that is consistent with the observation. The radius is set as 0.7~Mm according to the apparent width of the mini-filament. We first derive the reference equilibrium current-field intensity ($I_0$) according to Equation (7) in \cite{Titov2014}, denoted as $I_0 = 8.5 \times 10^{9}\;$A. With some tests and experiments, we select $I = 1.6 I_0$. As a result, the residual Lorentz force can directly trigger the eruption of the inserted flux rope.
Then, the toroidal flux $F$ is computed by Equation~(12) in \cite{Titov2018}.
The height of the null point, as measured by \cite{Cheng2023}, is remarkably low, consistent with our NLFFF extrapolation results, being only about 1-2~Mm above the photosphere. This suggests that the flux rope resides in the chromosphere, which does not conform to a force-free condition. As a result, we adopt an unstable flux rope resembling the eruptive mini-filament to initiate the spiral jet.

Initial 3D magnetic field in our MHD simulations is shown in Figure~\ref{figa1}. Then, we run the numerical simulation to derive the magnetic field evolution. The governing equations are as follows:

\begin{align}
    \frac{\partial\rho}{\partial t} + \nabla \cdot (\rho \boldsymbol{v}) &= 0 \\
    \frac{\partial(\rho\boldsymbol{v})}{\partial t} + 
    \nabla \cdot (\rho\boldsymbol{v}\boldsymbol{v} + p\boldsymbol{I} - \frac{\boldsymbol{BB}}{\mu_0}) 
    &= \rho \boldsymbol{g}\\
    \frac{\partial e_{int}}{\partial t} + \nabla \cdot (\boldsymbol{v}e_{int}) &= - p \nabla \cdot
    \boldsymbol{v}\\
    \frac{\partial \boldsymbol{B}}{\partial t} + 
    \nabla \cdot (\boldsymbol{vB - Bv}) &= 0
\end{align}
where $\rho$ is density, $\boldsymbol{v}$ is the velocity, $\boldsymbol{B}$ is the magnetic field, $e_{int} = p/(\gamma - 1)$ is the internal energy, $\boldsymbol{g}=-ge_{z}$ is the gravity acceleration. The 3D MHD equations are numerically solved with MPI-AMRVAC. The computational domain is 58.4 $\times$ 58.4 $\times$ 30\;Mm$^3$, resolved by a uniform grid with 300 $\times$ 300 $\times$ 600 cells. This setup yields a horizontal resolution of 195 km, maintaining the high resolution of the magnetogram, and a vertical resolution of 50 km, sufficient to accurately resolve the flux rope structure of the mini-filament in observations. The cadence of saved data in our simulation is $t_0 = 0.3 \tau$, where $\tau \approx 86 s$ is the Alfv$\acute{e}$n time.

\begin{figure*}[!ht]
    \centering
    \includegraphics[width=19cm]{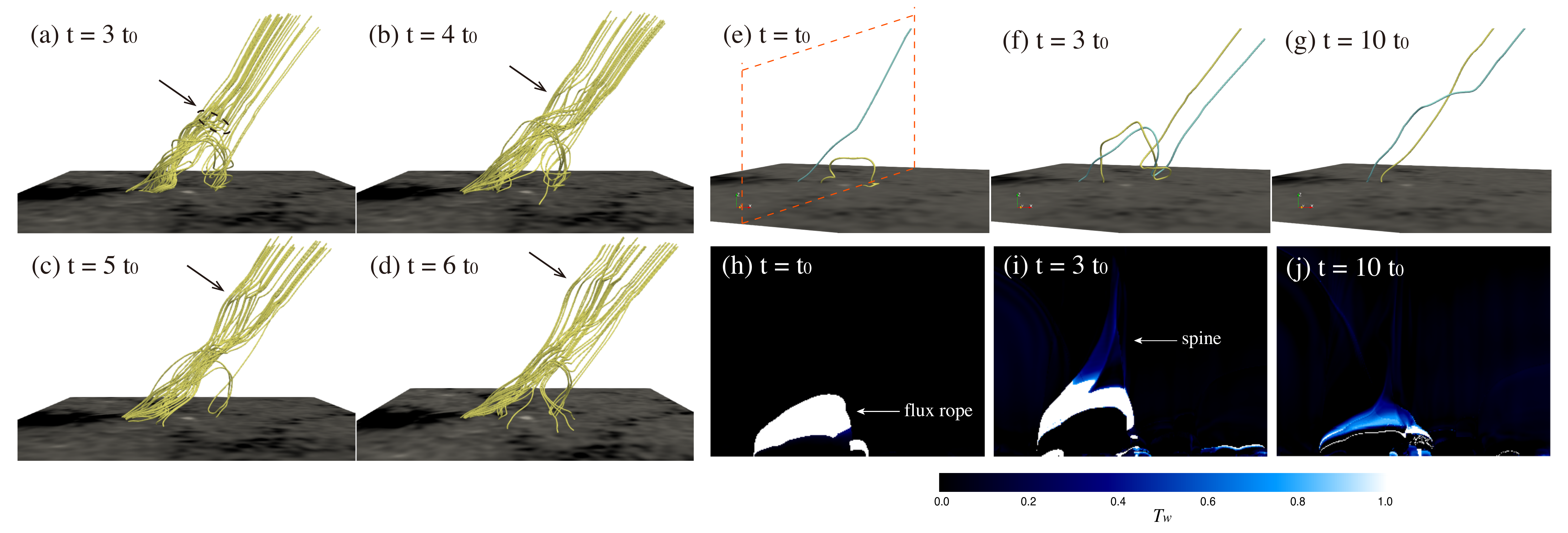}
    \caption{Twist transfer and propagation after the reconnection. (a) - (d): Evolution of magnetic field lines near the outer spine. Arrows point to the curvatures of the traced field lines. The ellipses in panel (g) indicate the plane to calculate the averaged mass flux in Figure \ref{figa3}. (e) - (g): Evolution of the representative filed line around the outer spine (blue) and the representative flux rope field line (yellow). (h) - (j): Evolution of the distribution of the twist number measured in turns in y-z plane outlined in panel (a).  } 
    \label{fig3}
\end{figure*}

\section{Results}
\label{results}

Here, we show the simulation results which present similarities with observations in Figure~\ref{fig1}(b) and \ref{fig1}(c). The orientation of the simulated spine (open field lines) is found to align closely with those of the observed jet outer spine. Note that the open field lines in our simulation correspond to closed field lines that connect to the distant polarities are not included within our computational domain, thus leading to the spine appearing as open field lines and being less inclined compared to the closed structures observed. For better visualization, the magnetic field lines are colored blue to red, indicating the transition from cool to warm plasma. This allows comparison with the observed dark and bright features. 

Before the eruption, the twisted flux rope resembling the mini-filament consisted of cool plasma and resided underneath the warm dome, similar to the observations. As the impulsive phase begins, the flux rope expands and ascends towards the overlying dome, eventually interacting and merging with it, releasing cool material to the spine (Figure \ref{fig1}(b2) and \ref{fig1}(c2)). The propagation speed of the erupting cool material is approximately 155 km s$^{-1}$, which is consistent with the speed measured in observations (See Figure \ref{figa2} for details). During this process, the erupting flux rope also expands and presents an untwisting motion. At the late stage of the simulation, the interaction between the flux rope and the dome persists, with cool material from the rope being continuously released and propagating along the outer spine (Figure \ref{fig1} (b3)(c3)). The side-view snapshot also illustrates that the main body of the flux rope below the dome has been completely dissipated. The highly twisted field lines appear above the dome, and cool materials are observed moving along the outer spine. The erupted cool flux shows a clear rotation motion, resembling the spiral motion of the observed dark threads accompanying the tiny jet. The strong resemblance between the simulation and observations indicates that our MHD model based on the SO/PHI high-resolution data basically reproduces the evolution of the observed tiny spiral jet.

A new result from the analysis of our MHD simulation is that an external reconnection is identified to take place in a curved current sheet as formed at the interaction region between the erupting flux rope and the dome, as shown in Figure~\ref{fig2}. The velocity vectors are presented by green and pink arrows, representing the positive and negative of $v_x$. One can see that these arrows point in opposite directions, indicating shear motions in the field lines, which suggests that the reconnection takes place between the twisted field and the ambient open field. Specifically, the velocities of upflows range from 60 km s$^{-1}$ to 190 km s$^{-1}$, consistent with the velocities of jets measured in observations \citep{Cheng2023, Shen2021}. Moreover, the lower section of the open field presents obvious dips (pointed out by green arrows in Figure \ref{fig2}(b)), indicating that they have just experienced a reconnection and started to be separated from the reconnection region. The increase of the temperature at magnetic dips of reconnected field lines, compared to surrounding regions, further infers local plasma heating. Given the limitation of the energy equation we adopted, the heating is likely mainly due to the compression driven by reconnection-induced flows. 
We image the curved current sheet in Figure \ref{fig2}(c) by displaying an iso-surface of $J/B = 0.5$ (1/$\Delta$) that best shows the region of enhanced currents. It denotes that the reconnection region is a curved 3D current sheet formed at the interface between the rising flux rope and the dome, similar to the theoretical models of spiral jets conceived in previous ideal MHD studies \citep{Masson2009}. It should be emphasized that it would be impossible to determine the location and 3D geometry of the reconnection for such a tiny-scale spiral jet without the high-resolution SO/PHI-HRT data constraining the numerical model. 

There is a clear transfer and propagation of magnetic twist following the external reconnection as illustrated in Figure \ref{fig3}. Initially, the twist is only confined within the flux rope, with the field lines near the outer spine showing no signs of twist. Figure \ref{fig3} (a) shows the configuration when the reconnection starts. One can see that the transfer of twist from the flux rope field line to the outer spine is evident. Subsequent snapshots depict the twist carried by the curved field lines and propagating towards the higher corona (Figure \ref{fig3} (b) - (d)).
We also estimate the propagation speed of the twist to be about 230 km s$^{-1}$. This speed is comparable to the local Alfv$\acute{e}$n speed, which is estimated to be approximately 300 km s$^{-1}$ based on a magnetic field strength of $B \sim 6\ G$, and the plasma density of $\rho \sim 2 \times 10^{-15} g\ cm^{-3}$ from our simulation data. This suggests that the reconnection between the twisted flux rope and the fan-spine launches a nonlinear torsional Alfv$\acute{e}$n wave that might transport energy and accelerate plasma along the outer spine \citep{Shibata1986, Pariat2009, Wyper2018, Kohutova2020}. 
We select two representative field lines, one around the outer spine and one within the flux rope, to illustrate the evolution of twist transfer (Figure \ref{fig3}(e)-(g)). Along with the evolution of the calculated twist number (Figure \ref{fig3} (h) - (j)), $T_w$, this clearly demonstrates how the twist is carried by the open field lines, while the flux rope field lines become untwisted because of the release of twist. The effective transfer of the twist implies that such tiny spiral jets could serve as a source of, at least, some switchbacks \citep{Bale2019, Sterling2020}.

\section{Summary and Discussion}
\label{summary}

Thanks to the high spatial resolution magnetogram provided by PHI-HRT, we are, for the first time, able to perform a data-constrained MHD simulation to study the spiral jet's generation mechanism from a tiny-scale fan-spine configuration. Our simulation successfully reproduces the spiral jet's observational features, including its propagation along the outer spine and rolling of associated dark and bright threads.

Coronal jets were proposed to be the miniature version of large-scale eruptions in previous observations \citep{Moore2010, Sterling2015} and ideal models \citep{Wyper2018}. For large-scale eruptions, similar data-constrained/driven models found that the unstable flux rope is an indispensable component for a full eruption \citep[e.g.,][]{Kliem2013}. 
However, these results have not been justified for small-scale jets, primarily due to the limitation of the spatial resolution of the bottom boundary magnetograms they used. By utilizing higher-resolution magnetic field data (195 km per pixel), our data-constrained model effectively resolves the eruption of a tiny-scale flux rope eruption spanning a scale of ~10 Mm that is highly comparable with EUI observations. In particular, our MHD model confirms the external reconnection between the flux rope and the ambient field and even discloses its detailed processes, similar to the results of \cite{Zhu2023} although they took advantage of zero-beta MHD simulation. On the other hand, the transfer of magnetic twist is previously suggested to occur in large-scale eruptions (e.g., \citealt{Masson2013, GuoJ2024}) and ideal MHD models \citep{Pariat2009, Wyper2017}. We further confirm here that the twist initially contained in the mini-flux rope can be released into the outer open fields through external reconnection. Additionally, because of the inclusion of the energy equation, despite only considering the adiabatic process, we also reproduce the transfer of cool materials along with that of the twist, which cannot be derived in zero-beta MHD simulations.

In previous jet models, the importance of external reconnection has already been highlighted \citep{Shibata1992, Moore2010, Sterling2015, Wyper2017}. Here, we further determine that the external magnetic reconnection occurs within a 3D curved current sheet, most likely collapsed from the pre-existing 3D magnetic null. We also find that such an external reconnection is the key to transferring the cool material and twist within the flux rope to the open flux close to the outer spine. As a matter of fact, numerous observational evidence for external reconnection such as small-scale blobs have been revealed by recent high-resolution imaging observations \citep{Li2023, Cheng2023}. 

There is one remarkable distinction between the external reconnection we discussed here and the breakout reconnection \citep{Wyper2017, Wyper2018}, which refers to the reconnection between the constraining field above the flux rope and the background field. In contrast, external reconnection occurs directly between the erupting flux rope and the background field, without involving a breakout process. Due to such an external reconnection, an erupting flux rope is even more easily confined due to the decrease of upward loop force, thus resulting in a failed eruption \citep{Chen2023}.

In addition, as our model represents an initial step in addressing the jet's configuration, it does not focus too much on the thermal evolution and therefore does not contain heat conduction and radiation processes in the energy equation. Consequently, the thermal properties obtained from this numerical model are not quantitatively precise, as the heating is driven by plasma compression ($-p\nabla \cdot \boldsymbol{v}$), which predominantly occurs near the current-sheet regions. Since key numerical heating terms, such as Joule heating and viscous heating, are omitted in this model, the plasma cannot reach the temperature values in observations \citep{Cheng2023}. Nevertheless, the temperature changes associated with the twist transfer are evident in Figure~\ref{fig2}. In future work, we will focus on improving the representation of the heating process by incorporating more comprehensive source terms.

\begin{acknowledgements}
We thank the referee who helped improve the manuscript. Solar Orbiter is a space mission of international collaboration between ESA and NASA, operated by ESA. We are grateful to the ESA SOC and MOC teams for their support. The German contribution to SO/PHI is funded by the BMWi through DLR and by MPG central funds. The Spanish contribution is funded by AEI/MCIN/10.13039/501100011033/ and European Union “NextGenerationEU”/PRTR” (RTI2018-096886-C5,  PID2021-125325OB-C5,  PCI2022-135009-2, PCI2022-135029-2) and ERDF “A way of making Europe”; “Center of Excellence Severo Ochoa” awards to IAA-CSIC (SEV-2017-0709, CEX2021-001131-S); and a Ramón y Cajal fellowship awarded to DOS. The French contribution is funded by CNES. The EUI instrument was built by CSL, IAS, MPS, MSSL/UCL, PMOD/WRC, ROB, LCF/IO with funding from the Belgian Federal Science Policy Office (BELSPO/PRODEX PEA 4000134088, 4000112292, 4000136424, and 4000134474); the Centre National d’Etudes Spatiales (CNES); the UK Space Agency (UKSA); the Bundesministerium für Wirtschaft und Energie (BMWi) through the Deutsches Zentrum für Luft- und Raumfahrt (DLR); and the Swiss Space Office (SSO). Z.F.L., X.C., and M.D.D. are funded by the Fundamental Research Funds for the Central Universities under grant 2024300348, NSFC grant 12127901, and National Key R\&D Program of China under grants 2021YFA1600504 and 2022YFF0503001. The simulations in this paper were performed in the cluster system of the High Performance Computing Center (HPCC) of Nanjing University. L.P.C. gratefully acknowledges funding by the European Union (ERC, ORIGIN, 101039844). J.H.G is supported by the China National Postdoctoral Program for Innovative Talents fellowship under Grant Number BX20240159. However, the views and opinions expressed are those of the author(s) only and do not necessarily reflect those of the European Union or the European Research Council. Neither the European Union nor the granting authority can be held responsible for them. SP acknowledges support from the projects C16/24/010 C1 project Internal Funds KU Leuven), G0B5823N and G002523N (WEAVE) (FWO-Vlaanderen), and 4000145223 (SIDC Data Exploitation (SIDEX2), ESA Prodex). We also thank Gherardo Valori for his assistance in processing the SO/PHI-HRT data.
\end{acknowledgements}

\newpage

\bibliography{bibtex}{}

\begin{thebibliography}{58}
\expandafter\ifx\csname natexlab\endcsname\relax\def\natexlab#1{#1}\fi

\bibitem[{{Bale} {et~al.}(2019){Bale}, {Badman}, {Bonnell}, {Bowen}, {Burgess}, {Case}, {Cattell}, {Chandran}, {Chaston}, {Chen}, {Drake}, {de Wit}, {Eastwood}, {Ergun}, {Farrell}, {Fong}, {Goetz}, {Goldstein}, {Goodrich}, {Harvey}, {Horbury}, {Howes}, {Kasper}, {Kellogg}, {Klimchuk}, {Korreck}, {Krasnoselskikh}, {Krucker}, {Laker}, {Larson}, {MacDowall}, {Maksimovic}, {Malaspina}, {Martinez-Oliveros}, {McComas}, {Meyer-Vernet}, {Moncuquet}, {Mozer}, {Phan}, {Pulupa}, {Raouafi}, {Salem}, {Stansby}, {Stevens}, {Szabo}, {Velli}, {Woolley}, \& {Wygant}}]{Bale2019}
{Bale}, S.~D., {Badman}, S.~T., {Bonnell}, J.~W., {et~al.} 2019, \nat, 576, 237

\bibitem[{{Calchetti} {et~al.}(2023){Calchetti}, {Stangalini}, {Jafarzadeh}, {Valori}, {Albert}, {Albelo Jorge}, {Alvarez-Herrero}, {Appourchaux}, {Balaguer Jim{\'e}nez}, {Bellot Rubio}, {Blanco Rodr{\'\i}guez}, {Feller}, {Gandorfer}, {Germerott}, {Gizon}, {Guerrero}, {Gutierrez-Marques}, {Hirzberger}, {Kahil}, {Kolleck}, {Korpi-Lagg}, {Moreno Vacas}, {Orozco Su{\'a}rez}, {P{\'e}rez-Grande}, {Sanchis Kilders}, {Schou}, {Sch{\"u}hle}, {Sinjan}, {Solanki}, {Staub}, {Strecker}, {del Toro Iniesta}, {Volkmer}, \& {Woch}}]{Calchetti2023}
{Calchetti}, D., {Stangalini}, M., {Jafarzadeh}, S., {et~al.} 2023, \aap, 674, A109

\bibitem[{{Chen} {et~al.}(2023){Chen}, {Cheng}, {Kliem}, \& {Ding}}]{Chen2023}
{Chen}, J., {Cheng}, X., {Kliem}, B., \& {Ding}, M. 2023, \apjl, 951, L35

\bibitem[{{Cheng} {et~al.}(2023){Cheng}, {Priest}, {Li}, {Chen}, {Aulanier}, {Chitta}, {Wang}, {Peter}, {Zhu}, {Xing}, {Ding}, {Solanki}, {Berghmans}, {Teriaca}, {Aznar Cuadrado}, {Zhukov}, {Guo}, {Long}, {Harra}, {Smith}, {Rodriguez}, {Verbeeck}, {Barczynski}, \& {Parenti}}]{Cheng2023}
{Cheng}, X., {Priest}, E.~R., {Li}, H.~T., {et~al.} 2023, Nature Communications, 14, 2107

\bibitem[{{Cheung} {et~al.}(2015){Cheung}, {De Pontieu}, {Tarbell}, {Fu}, {Tian}, {Testa}, {Reeves}, {Mart{\'\i}nez-Sykora}, {Boerner}, {W{\"u}lser}, {Lemen}, {Title}, {Hurlburt}, {Kleint}, {Kankelborg}, {Jaeggli}, {Golub}, {McKillop}, {Saar}, {Carlsson}, \& {Hansteen}}]{Cheung2015}
{Cheung}, M. C.~M., {De Pontieu}, B., {Tarbell}, T.~D., {et~al.} 2015, \apj, 801, 83

\bibitem[{{Chitta} {et~al.}(2022){Chitta}, {Peter}, {Parenti}, {Berghmans}, {Auch{\`e}re}, {Solanki}, {Aznar Cuadrado}, {Sch{\"u}hle}, {Teriaca}, {Mandal}, {Barczynski}, {Buchlin}, {Harra}, {Kraaikamp}, {Long}, {Rodriguez}, {Schwanitz}, {Smith}, {Verbeeck}, {Zhukov}, {Liu}, \& {Cheung}}]{Chitta2022}
{Chitta}, L.~P., {Peter}, H., {Parenti}, S., {et~al.} 2022, \aap, 667, A166

\bibitem[{{Chitta} {et~al.}(2023){Chitta}, {Zhukov}, {Berghmans}, {Peter}, {Parenti}, {Mandal}, {Aznar Cuadrado}, {Sch{\"u}hle}, {Teriaca}, {Auch{\`e}re}, {Barczynski}, {Buchlin}, {Harra}, {Kraaikamp}, {Long}, {Rodriguez}, {Schwanitz}, {Smith}, {Verbeeck}, \& {Seaton}}]{Chitta2023}
{Chitta}, L.~P., {Zhukov}, A.~N., {Berghmans}, D., {et~al.} 2023, Science, 381, 867

\bibitem[{{Gandorfer} {et~al.}(2018){Gandorfer}, {Grauf}, {Staub}, {Bischoff}, {Woch}, {Hirzberger}, {Solanki}, {{\'A}lvarez-Herrero}, {Garc{\'\i}a Parejo}, {Schmidt}, {Volkmer}, {Appourchaux}, \& {del Toro Iniesta}}]{Gandorfer2018}
{Gandorfer}, A., {Grauf}, B., {Staub}, J., {et~al.} 2018, in Society of Photo-Optical Instrumentation Engineers (SPIE) Conference Series, Vol. 10698, Space Telescopes and Instrumentation 2018: Optical, Infrared, and Millimeter Wave, ed. M.~{Lystrup}, H.~A. {MacEwen}, G.~G. {Fazio}, N.~{Batalha}, N.~{Siegler}, \& E.~C. {Tong}, 106984N

\bibitem[{{Guo} {et~al.}(2024){Guo}, {Linan}, {Poedts}, {Guo}, {Lani}, {Schmieder}, {Brchnelova}, {Perri}, {Baratashvili}, {Ni}, \& {Chen}}]{GuoJ2024}
{Guo}, J.~H., {Linan}, L., {Poedts}, S., {et~al.} 2024, \aap, 683, A54

\bibitem[{{Guo} {et~al.}(2023){Guo}, {Ni}, {Zhong}, {Guo}, {Xia}, {Li}, {Poedts}, {Schmieder}, \& {Chen}}]{Guo2023}
{Guo}, J.~H., {Ni}, Y.~W., {Zhong}, Z., {et~al.} 2023, \apjs, 266, 3

\bibitem[{{Hoeksema} {et~al.}(2014){Hoeksema}, {Liu}, {Hayashi}, {Sun}, {Schou}, {Couvidat}, {Norton}, {Bobra}, {Centeno}, {Leka}, {Barnes}, \& {Turmon}}]{Hoeksema2014}
{Hoeksema}, J.~T., {Liu}, Y., {Hayashi}, K., {et~al.} 2014, \solphys, 289, 3483

\bibitem[{{Innes} {et~al.}(1997){Innes}, {Inhester}, {Axford}, \& {Wilhelm}}]{Innes1997}
{Innes}, D.~E., {Inhester}, B., {Axford}, W.~I., \& {Wilhelm}, K. 1997, \nat, 386, 811

\bibitem[{{Jiang} {et~al.}(2016){Jiang}, {Wu}, {Feng}, \& {Hu}}]{Jiang2016}
{Jiang}, C., {Wu}, S.~T., {Feng}, X., \& {Hu}, Q. 2016, Nature Communications, 7, 11522

\bibitem[{{Joshi} {et~al.}(2020){Joshi}, {Wang}, {Chandra}, {Zhang}, {Liu}, \& {Li}}]{Joshi2020b}
{Joshi}, R., {Wang}, Y., {Chandra}, R., {et~al.} 2020, \apj, 901, 94

\bibitem[{{Kahil} {et~al.}(2022){Kahil}, {Hirzberger}, {Solanki}, {Chitta}, {Peter}, {Auch{\`e}re}, {Sinjan}, {Orozco Su{\'a}rez}, {Albert}, {Albelo Jorge}, {Appourchaux}, {Alvarez-Herrero}, {Blanco Rodr{\'\i}guez}, {Gandorfer}, {Germerott}, {Guerrero}, {Guti{\'e}rrez M{\'a}rquez}, {Kolleck}, {del Toro Iniesta}, {Volkmer}, {Woch}, {Fiethe}, {G{\'o}mez Cama}, {P{\'e}rez-Grande}, {Sanchis Kilders}, {Balaguer Jim{\'e}nez}, {Bellot Rubio}, {Calchetti}, {Carmona}, {Deutsch}, {Fern{\'a}ndez-Rico}, {Fern{\'a}ndez-Medina}, {Garc{\'\i}a Parejo}, {Gasent-Blesa}, {Gizon}, {Grauf}, {Heerlein}, {Lagg}, {Lange}, {L{\'o}pez Jim{\'e}nez}, {Maue}, {Meller}, {Michalik}, {Moreno Vacas}, {M{\"u}ller}, {Nakai}, {Schmidt}, {Schou}, {Sch{\"u}hle}, {Staub}, {Strecker}, {Torralbo}, {Valori}, {Aznar Cuadrado}, {Teriaca}, {Berghmans}, {Verbeeck}, {Kraaikamp}, \& {Gissot}}]{Kahil2022}
{Kahil}, F., {Hirzberger}, J., {Solanki}, S.~K., {et~al.} 2022, \aap, 660, A143

\bibitem[{{Keppens} {et~al.}(2012){Keppens}, {Meliani}, {van Marle}, {Delmont}, {Vlasis}, \& {van der Holst}}]{Keppens2012}
{Keppens}, R., {Meliani}, Z., {van Marle}, A.~J., {et~al.} 2012, Journal of Computational Physics, 231, 718

\bibitem[{{Keppens} {et~al.}(2003){Keppens}, {Nool}, {T{\'o}th}, \& {Goedbloed}}]{Keppens2003}
{Keppens}, R., {Nool}, M., {T{\'o}th}, G., \& {Goedbloed}, J.~P. 2003, Computer Physics Communications, 153, 317

\bibitem[{Kliem {et~al.}(2013)Kliem, Su, van Ballegooijen, \& DeLuca}]{Kliem2013}
Kliem, B., Su, Y.~N., van Ballegooijen, A.~A., \& DeLuca, E.~E. 2013, The Astrophysical Journal, 779, 129

\bibitem[{{Kohutova} {et~al.}(2020){Kohutova}, {Verwichte}, \& {Froment}}]{Kohutova2020}
{Kohutova}, P., {Verwichte}, E., \& {Froment}, C. 2020, \aap, 633, L6

\bibitem[{{Lemen} {et~al.}(2012){Lemen}, {Title}, {Akin}, {Boerner}, {Chou}, {Drake}, {Duncan}, {Edwards}, {Friedlaender}, {Heyman}, {Hurlburt}, {Katz}, {Kushner}, {Levay}, {Lindgren}, {Mathur}, {McFeaters}, {Mitchell}, {Rehse}, {Schrijver}, {Springer}, {Stern}, {Tarbell}, {Wuelser}, {Wolfson}, {Yanari}, {Bookbinder}, {Cheimets}, {Caldwell}, {Deluca}, {Gates}, {Golub}, {Park}, {Podgorski}, {Bush}, {Scherrer}, {Gummin}, {Smith}, {Auker}, {Jerram}, {Pool}, {Soufli}, {Windt}, {Beardsley}, {Clapp}, {Lang}, \& {Waltham}}]{Lemen2012}
{Lemen}, J.~R., {Title}, A.~M., {Akin}, D.~J., {et~al.} 2012, \solphys, 275, 17

\bibitem[{{Li} {et~al.}(2023){Li}, {Cheng}, {Ding}, {Chitta}, {Peter}, {Berghmans}, {Smith}, {Auch{\`e}re}, {Parenti}, {Barczynski}, {Harra}, {Sch{\"u}hle}, {Buchlin}, {Verbeeck}, {Aznar Cuadrado}, {Zhukov}, {Long}, {Teriaca}, \& {Rodriguez}}]{Li2023}
{Li}, Z.~F., {Cheng}, X., {Ding}, M.~D., {et~al.} 2023, \aap, 673, A83

\bibitem[{{Liu} {et~al.}(2015){Liu}, {Wang}, {Shen}, {Liu}, {Pan}, \& {Wang}}]{Liujj2015}
{Liu}, J., {Wang}, Y., {Shen}, C., {et~al.} 2015, \apj, 813, 115

\bibitem[{Masson {et~al.}(2013)Masson, Antiochos, \& DeVore}]{Masson2013}
Masson, S., Antiochos, S.~K., \& DeVore, C.~R. 2013, The Astrophysical Journal, 771, 82

\bibitem[{{Masson} {et~al.}(2009){Masson}, {Pariat}, {Aulanier}, \& {Schrijver}}]{Masson2009}
{Masson}, S., {Pariat}, E., {Aulanier}, G., \& {Schrijver}, C.~J. 2009, \apj, 700, 559

\bibitem[{{Metcalf}(1994)}]{Metcalf1994}
{Metcalf}, T.~R. 1994, \solphys, 155, 235

\bibitem[{{Metcalf} {et~al.}(2006){Metcalf}, {Leka}, {Barnes}, {Lites}, {Georgoulis}, {Pevtsov}, {Balasubramaniam}, {Gary}, {Jing}, {Li}, {Liu}, {Wang}, {Abramenko}, {Yurchyshyn}, \& {Moon}}]{Metcalf2006}
{Metcalf}, T.~R., {Leka}, K.~D., {Barnes}, G., {et~al.} 2006, \solphys, 237, 267

\bibitem[{{Moore} {et~al.}(2010){Moore}, {Cirtain}, {Sterling}, \& {Falconer}}]{Moore2010}
{Moore}, R.~L., {Cirtain}, J.~W., {Sterling}, A.~C., \& {Falconer}, D.~A. 2010, \apj, 720, 757

\bibitem[{Moore {et~al.}(2011)Moore, Sterling, Cirtain, \& Falconer}]{Moore2011}
Moore, R.~L., Sterling, A.~C., Cirtain, J.~W., \& Falconer, D.~A. 2011, The Astrophysical Journal Letters, 731, L18

\bibitem[{{M{\"u}ller} {et~al.}(2020){M{\"u}ller}, {St. Cyr}, {Zouganelis}, {Gilbert}, {Marsden}, {Nieves-Chinchilla}, {Antonucci}, {Auch{\`e}re}, {Berghmans}, {Horbury}, {Howard}, {Krucker}, {Maksimovic}, {Owen}, {Rochus}, {Rodriguez-Pacheco}, {Romoli}, {Solanki}, {Bruno}, {Carlsson}, {Fludra}, {Harra}, {Hassler}, {Livi}, {Louarn}, {Peter}, {Sch{\"u}hle}, {Teriaca}, {del Toro Iniesta}, {Wimmer-Schweingruber}, {Marsch}, {Velli}, {De Groof}, {Walsh}, \& {Williams}}]{Muller2020}
{M{\"u}ller}, D., {St. Cyr}, O.~C., {Zouganelis}, I., {et~al.} 2020, \aap, 642, A1

\bibitem[{{Nayak} {et~al.}(2019){Nayak}, {Bhattacharyya}, {Prasad}, {Hu}, {Kumar}, \& {Joshi}}]{Nayak2019}
{Nayak}, S.~S., {Bhattacharyya}, R., {Prasad}, A., {et~al.} 2019, \apj, 875, 10

\bibitem[{{Pariat} {et~al.}(2009){Pariat}, {Antiochos}, \& {DeVore}}]{Pariat2009}
{Pariat}, E., {Antiochos}, S.~K., \& {DeVore}, C.~R. 2009, \apj, 691, 61

\bibitem[{{Pariat} {et~al.}(2010){Pariat}, {Antiochos}, \& {DeVore}}]{Pariat2010}
{Pariat}, E., {Antiochos}, S.~K., \& {DeVore}, C.~R. 2010, \apj, 714, 1762

\bibitem[{{Pesnell} {et~al.}(2012){Pesnell}, {Thompson}, \& {Chamberlin}}]{Pesnell2012}
{Pesnell}, W.~D., {Thompson}, B.~J., \& {Chamberlin}, P.~C. 2012, \solphys, 275, 3

\bibitem[{{Priest} \& {Syntelis}(2021)}]{Priest2021}
{Priest}, E.~R. \& {Syntelis}, P. 2021, \aap, 647, A31

\bibitem[{{Raouafi} {et~al.}(2016){Raouafi}, {Patsourakos}, {Pariat}, {Young}, {Sterling}, {Savcheva}, {Shimojo}, {Moreno-Insertis}, {DeVore}, {Archontis}, {T{\"o}r{\"o}k}, {Mason}, {Curdt}, {Meyer}, {Dalmasse}, \& {Matsui}}]{Raouafi2016}
{Raouafi}, N.~E., {Patsourakos}, S., {Pariat}, E., {et~al.} 2016, \ssr, 201, 1

\bibitem[{{Rochus} {et~al.}(2020){Rochus}, {Auch{\`e}re}, {Berghmans}, {Harra}, {Schmutz}, {Sch{\"u}hle}, {Addison}, {Appourchaux}, {Aznar Cuadrado}, {Baker}, {Barbay}, {Bates}, {BenMoussa}, {Bergmann}, {Beurthe}, {Borgo}, {Bonte}, {Bouzit}, {Bradley}, {B{\"u}chel}, {Buchlin}, {B{\"u}chner}, {Cab{\'e}}, {Cadiergues}, {Chaigneau}, {Chares}, {Choque Cortez}, {Coker}, {Condamin}, {Coumar}, {Curdt}, {Cutler}, {Davies}, {Davison}, {Defise}, {Del Zanna}, {Delmotte}, {Delouille}, {Dolla}, {Dumesnil}, {D{\"u}rig}, {Enge}, {Fran{\c{c}}ois}, {Fourmond}, {Gillis}, {Giordanengo}, {Gissot}, {Green}, {Guerreiro}, {Guilbaud}, {Gyo}, {Haberreiter}, {Hafiz}, {Hailey}, {Halain}, {Hansotte}, {Hecquet}, {Heerlein}, {Hellin}, {Hemsley}, {Hermans}, {Hervier}, {Hochedez}, {Houbrechts}, {Ihsan}, {Jacques}, {J{\'e}r{\^o}me}, {Jones}, {Kahle}, {Kennedy}, {Klaproth}, {Kolleck}, {Koller}, {Kotsialos}, {Kraaikamp}, {Langer}, {Lawrenson}, {Le Clech'}, {Lenaerts}, {Liebecq}, {Linder}, {Long}, {Mampaey}, {Markiewicz-Innes}, {Marquet},
  {Marsch}, {Matthews}, {Mazy}, {Mazzoli}, {Meining}, {Meltchakov}, {Mercier}, {Meyer}, {Monecke}, {Monfort}, {Morinaud}, {Moron}, {Mountney}, {M{\"u}ller}, {Nicula}, {Parenti}, {Peter}, {Pfiffner}, {Philippon}, {Phillips}, {Plesseria}, {Pylyser}, {Rabecki}, {Ravet-Krill}, {Rebellato}, {Renotte}, {Rodriguez}, {Roose}, {Rosin}, {Rossi}, {Roth}, {Rouesnel}, {Roulliay}, {Rousseau}, {Ruane}, {Scanlan}, {Schlatter}, {Seaton}, {Silliman}, {Smit}, {Smith}, {Solanki}, {Spescha}, {Spencer}, {Stegen}, {Stockman}, {Szwec}, {Tamiatto}, {Tandy}, {Teriaca}, {Theobald}, {Tychon}, {van Driel-Gesztelyi}, {Verbeeck}, {Vial}, {Werner}, {West}, {Westwood}, {Wiegelmann}, {Willis}, {Winter}, {Zerr}, {Zhang}, \& {Zhukov}}]{Rochus2020}
{Rochus}, P., {Auch{\`e}re}, F., {Berghmans}, D., {et~al.} 2020, \aap, 642, A8

\bibitem[{{Scherrer} {et~al.}(2012){Scherrer}, {Schou}, {Bush}, {Kosovichev}, {Bogart}, {Hoeksema}, {Liu}, {Duvall}, {Zhao}, {Title}, {Schrijver}, {Tarbell}, \& {Tomczyk}}]{Scherrer2012}
{Scherrer}, P.~H., {Schou}, J., {Bush}, R.~I., {et~al.} 2012, \solphys, 275, 207

\bibitem[{{Schmieder} {et~al.}(2014){Schmieder}, {Archontis}, \& {Pariat}}]{Schmieder2014}
{Schmieder}, B., {Archontis}, V., \& {Pariat}, E. 2014, \ssr, 186, 227

\bibitem[{{Shen}(2021)}]{Shen2021}
{Shen}, Y. 2021, Proceedings of the Royal Society of London Series A, 477, 217

\bibitem[{{Shibata} {et~al.}(1992){Shibata}, {Ishido}, {Acton}, {Strong}, {Hirayama}, {Uchida}, {McAllister}, {Matsumoto}, {Tsuneta}, {Shimizu}, {Hara}, {Sakurai}, {Ichimoto}, {Nishino}, \& {Ogawara}}]{Shibata1992}
{Shibata}, K., {Ishido}, Y., {Acton}, L.~W., {et~al.} 1992, \pasj, 44, L173

\bibitem[{{Shibata} {et~al.}(2007){Shibata}, {Nakamura}, {Matsumoto}, {Otsuji}, {Okamoto}, {Nishizuka}, {Kawate}, {Watanabe}, {Nagata}, {UeNo}, {Kitai}, {Nozawa}, {Tsuneta}, {Suematsu}, {Ichimoto}, {Shimizu}, {Katsukawa}, {Tarbell}, {Berger}, {Lites}, {Shine}, \& {Title}}]{Shibata2007}
{Shibata}, K., {Nakamura}, T., {Matsumoto}, T., {et~al.} 2007, Science, 318, 1591

\bibitem[{{Shibata} \& {Uchida}(1986)}]{Shibata1986}
{Shibata}, K. \& {Uchida}, Y. 1986, \solphys, 103, 299

\bibitem[{{Shimojo} \& {Shibata}(2000)}]{Shimojo2000}
{Shimojo}, M. \& {Shibata}, K. 2000, Advances in Space Research, 26, 449

\bibitem[{{Sinjan} {et~al.}(2022){Sinjan}, {Calchetti}, {Hirzberger}, {Orozco Su{\'a}rez}, {Albert}, {Albelo Jorge}, {Appourchaux}, {Alvarez-Herrero}, {Blanco Rodr{\'\i}guez}, {Gandorfer}, {Germerott}, {Guerrero}, {Gutierrez Marquez}, {Kahil}, {Kolleck}, {Solanki}, {del Toro Iniesta}, {Volkmer}, {Woch}, {Fiethe}, {G{\'o}mez Cama}, {P{\'e}rez-Grande}, {Sanchis Kilders}, {Balaguer Jim{\'e}nez}, {Bellot Rubio}, {Carmona}, {Deutsch}, {Fernandez-Rico}, {Fern{\'a}ndez-Medina}, {Garc{\'\i}a Parejo}, {Gasent Blesa}, {Gizon}, {Grauf}, {Heerlein}, {Korpi-Lagg}, {Lange}, {L{\'o}pez Jim{\'e}nez}, {Maue}, {Meller}, {Michalik}, {Moreno Vacas}, {M{\"u}ller}, {Nakai}, {Schmidt}, {Schou}, {Sch{\"u}hle}, {Staub}, {Strecker}, {Torralbo}, \& {Valori}}]{Sinjan2022}
{Sinjan}, J., {Calchetti}, D., {Hirzberger}, J., {et~al.} 2022, in Society of Photo-Optical Instrumentation Engineers (SPIE) Conference Series, Vol. 12189, Software and Cyberinfrastructure for Astronomy VII, 121891J

\bibitem[{{Solanki} {et~al.}(2020){Solanki}, {del Toro Iniesta}, {Woch}, {Gandorfer}, {Hirzberger}, {Alvarez-Herrero}, {Appourchaux}, {Mart{\'\i}nez Pillet}, {P{\'e}rez-Grande}, {Sanchis Kilders}, {Schmidt}, {G{\'o}mez Cama}, {Michalik}, {Deutsch}, {Fernandez-Rico}, {Grauf}, {Gizon}, {Heerlein}, {Kolleck}, {Lagg}, {Meller}, {M{\"u}ller}, {Sch{\"u}hle}, {Staub}, {Albert}, {Alvarez Copano}, {Beckmann}, {Bischoff}, {Busse}, {Enge}, {Frahm}, {Germerott}, {Guerrero}, {L{\"o}ptien}, {Meierdierks}, {Oberdorfer}, {Papagiannaki}, {Ramanath}, {Schou}, {Werner}, {Yang}, {Zerr}, {Bergmann}, {Bochmann}, {Heinrichs}, {Meyer}, {Monecke}, {M{\"u}ller}, {Sperling}, {{\'A}lvarez Garc{\'\i}a}, {Aparicio}, {Balaguer Jim{\'e}nez}, {Bellot Rubio}, {Cobos Carracosa}, {Girela}, {Hern{\'a}ndez Exp{\'o}sito}, {Herranz}, {Labrousse}, {L{\'o}pez Jim{\'e}nez}, {Orozco Su{\'a}rez}, {Ramos}, {Barandiar{\'a}n}, {Bastide}, {Campuzano}, {Cebollero}, {D{\'a}vila}, {Fern{\'a}ndez-Medina}, {Garc{\'\i}a Parejo}, {Garranzo-Garc{\'\i}a}, {Laguna},
  {Mart{\'\i}n}, {Navarro}, {N{\'u}{\~n}ez Peral}, {Royo}, {S{\'a}nchez}, {Silva-L{\'o}pez}, {Vera}, {Villanueva}, {Fourmond}, {de Galarreta}, {Bouzit}, {Hervier}, {Le Clec'h}, {Szwec}, {Chaigneau}, {Buttice}, {Dominguez-Tagle}, {Philippon}, {Boumier}, {Le Cocguen}, {Baranjuk}, {Bell}, {Berkefeld}, {Baumgartner}, {Heidecke}, {Maue}, {Nakai}, {Scheiffelen}, {Sigwarth}, {Soltau}, {Volkmer}, {Blanco Rodr{\'\i}guez}, {Domingo}, {Ferreres Sabater}, {Gasent Blesa}, {Rodr{\'\i}guez Mart{\'\i}nez}, {Osorno Caudel}, {Bosch}, {Casas}, {Carmona}, {Herms}, {Roma}, {Alonso}, {G{\'o}mez-Sanjuan}, {Piqueras}, {Torralbo}, {Fiethe}, {Guan}, {Lange}, {Michel}, {Bonet}, {Fahmy}, {M{\"u}ller}, \& {Zouganelis}}]{Solanki2020}
{Solanki}, S.~K., {del Toro Iniesta}, J.~C., {Woch}, J., {et~al.} 2020, \aap, 642, A11

\bibitem[{{Sterling} {et~al.}(2015){Sterling}, {Moore}, {Falconer}, \& {Adams}}]{Sterling2015}
{Sterling}, A.~C., {Moore}, R.~L., {Falconer}, D.~A., \& {Adams}, M. 2015, \nat, 523, 437

\bibitem[{{Sterling} {et~al.}(2020){Sterling}, {Moore}, {Panesar}, \& {Samanta}}]{Sterling2020}
{Sterling}, A.~C., {Moore}, R.~L., {Panesar}, N.~K., \& {Samanta}, T. 2020, in Journal of Physics Conference Series, Vol. 1620, Journal of Physics Conference Series (IOP), 012020

\bibitem[{{Tian} {et~al.}(2018){Tian}, {Zhu}, {Peter}, {Zhao}, {Samanta}, \& {Chen}}]{Tian2018}
{Tian}, H., {Zhu}, X., {Peter}, H., {et~al.} 2018, \apj, 854, 174

\bibitem[{{Titov} {et~al.}(2018){Titov}, {Downs}, {Miki{\'c}}, {T{\"o}r{\"o}k}, {Linker}, \& {Caplan}}]{Titov2018}
{Titov}, V.~S., {Downs}, C., {Miki{\'c}}, Z., {et~al.} 2018, \apjl, 852, L21

\bibitem[{Titov {et~al.}(2014)Titov, Török, Mikic, \& Linker}]{Titov2014}
Titov, V.~S., Török, T., Mikic, Z., \& Linker, J.~A. 2014, The Astrophysical Journal, 790, 163

\bibitem[{{Wang} {et~al.}(1998){Wang}, {Sheeley}, {Socker}, {Howard}, {Brueckner}, {Michels}, {Moses}, {St. Cyr}, {Llebaria}, \& {Delaboudini{\`e}re}}]{Wang1998}
{Wang}, Y.~M., {Sheeley}, N.~R., J., {Socker}, D.~G., {et~al.} 1998, \apj, 508, 899

\bibitem[{{Wyper} {et~al.}(2017){Wyper}, {Antiochos}, \& {DeVore}}]{Wyper2017}
{Wyper}, P.~F., {Antiochos}, S.~K., \& {DeVore}, C.~R. 2017, \nat, 544, 452

\bibitem[{{Wyper} \& {DeVore}(2016)}]{Wyper2016}
{Wyper}, P.~F. \& {DeVore}, C.~R. 2016, \apj, 820, 77

\bibitem[{{Wyper} {et~al.}(2018){Wyper}, {DeVore}, \& {Antiochos}}]{Wyper2018}
{Wyper}, P.~F., {DeVore}, C.~R., \& {Antiochos}, S.~K. 2018, \apj, 852, 98

\bibitem[{{Wyper} {et~al.}(2019){Wyper}, {DeVore}, \& {Antiochos}}]{Wyper2019}
{Wyper}, P.~F., {DeVore}, C.~R., \& {Antiochos}, S.~K. 2019, \mnras, 490, 3679

\bibitem[{{Xia} {et~al.}(2018){Xia}, {Teunissen}, {El Mellah}, {Chan{\'e}}, \& {Keppens}}]{Xia2018}
{Xia}, C., {Teunissen}, J., {El Mellah}, I., {Chan{\'e}}, E., \& {Keppens}, R. 2018, \apjs, 234, 30

\bibitem[{{Yokoyama} \& {Shibata}(1996)}]{Yokoyama1996}
{Yokoyama}, T. \& {Shibata}, K. 1996, \pasj, 48, 353

\bibitem[{{Zhu} {et~al.}(2023){Zhu}, {Guo}, {Ding}, \& {Schmieder}}]{Zhu2023}
{Zhu}, J., {Guo}, Y., {Ding}, M., \& {Schmieder}, B. 2023, \apj, 949, 2

\end{thebibliography}
\bibliographystyle{aa}

\begin{appendix}

\section{Overview of the surface magnetic field}

\begin{figure}[!ht]
    \centering
    \includegraphics[width=6cm]{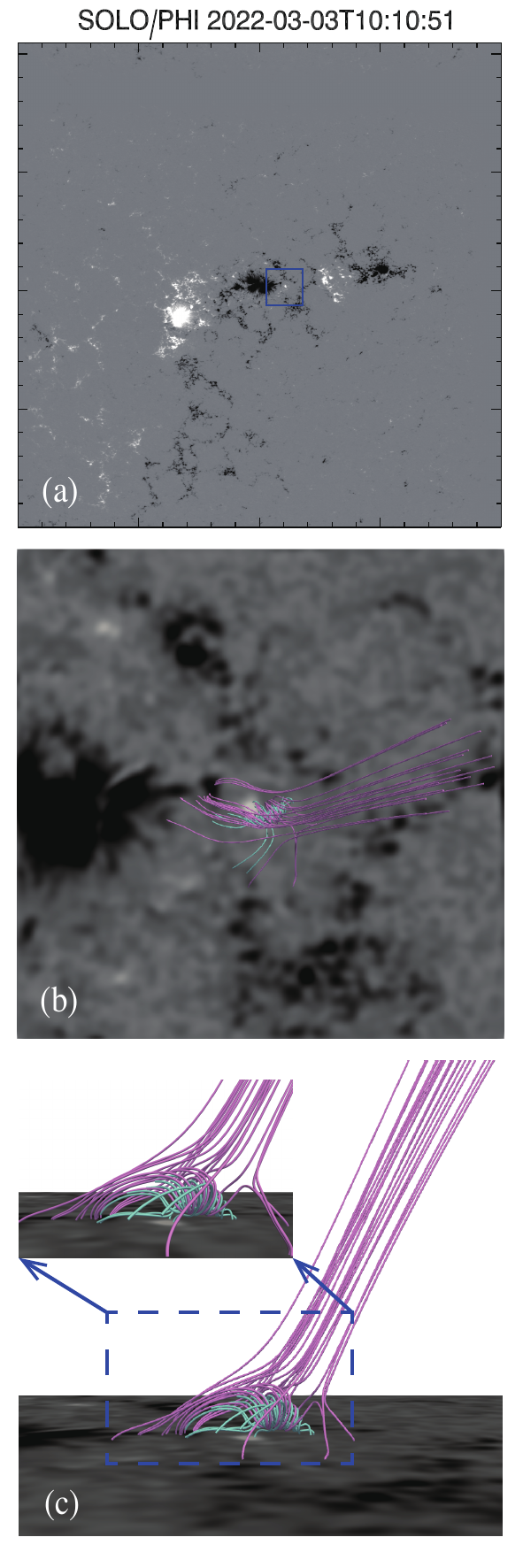}
    \caption{Overview of the surface magnetic field. (a) PHI line-of-sight (LOS) magnetogram showing the magnetic field distribution of the associated main active region before the eruption. (b) 3D magnetic field configuration, traced by the magnetic field lines, showing the initial condition of data-constrained 3D MHD simulation. The background image is the zoom-in of the PHI LOS magnetogram, as marked by a blue square in panel (a). (c) It is the same as panel (b) but for the side view and with a small panel more clearly showing the flux rope. } 
    \label{figa1}
\end{figure}

We present an overview of the surface magnetic field in Figure~\ref{figa1}(a) and initial 3D magnetic field of our MHD simulations in Figure~\ref{figa1}(b)(c). Specifically, this structure reveals a dome containing a 3D null, representing the fan separatrix surface of magnetic field lines enclosing the inserted flux rope and connecting towards the polarity outside the simulation box. This topology is consistent with the suggestion in \cite{Cheng2023}. 

\section{Propagation of cool material \label{app1}}

\begin{figure}[h]
    \begin{center}
    \includegraphics[width=0.5\textwidth,clip,trim=0cm 0cm 0cm 0cm]{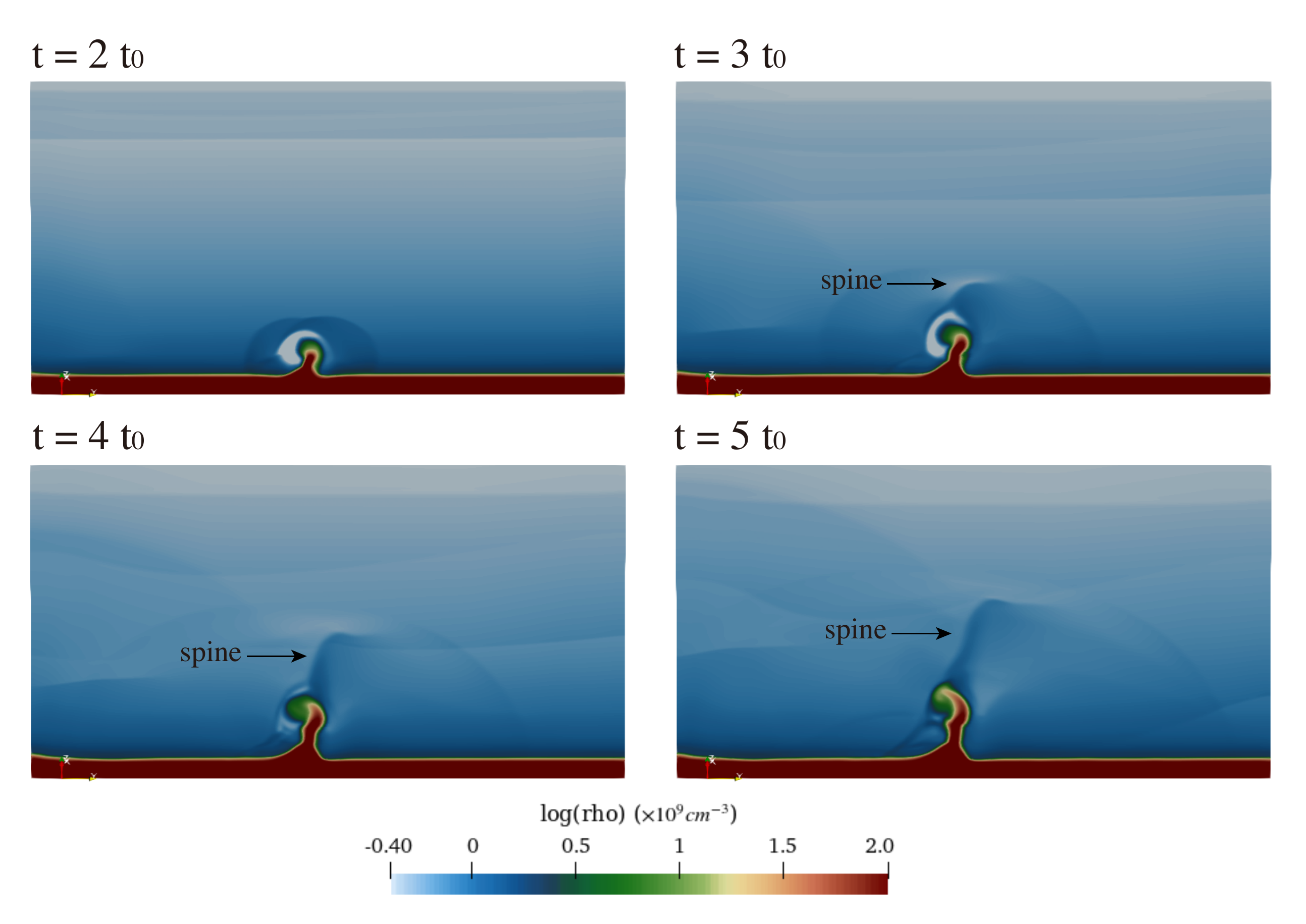}
    \caption{Distribution of density during the tiny jet showing the propagation of an erupting blob in the plane vertical to the spine.}
    \label{figa2}
    \end{center}
\end{figure}

Figure \ref{figa2} clearly illustrates the propagation of an erupting blob with dense plasma along the outer spine. Initially, the flux rope contains cool, dense plasma. After the reconnection, the dense plasma is released to the outer spine and subsequently propagates toward the upper corona.


\begin{figure}[!ht]
    \centering
    \includegraphics[width=8cm]{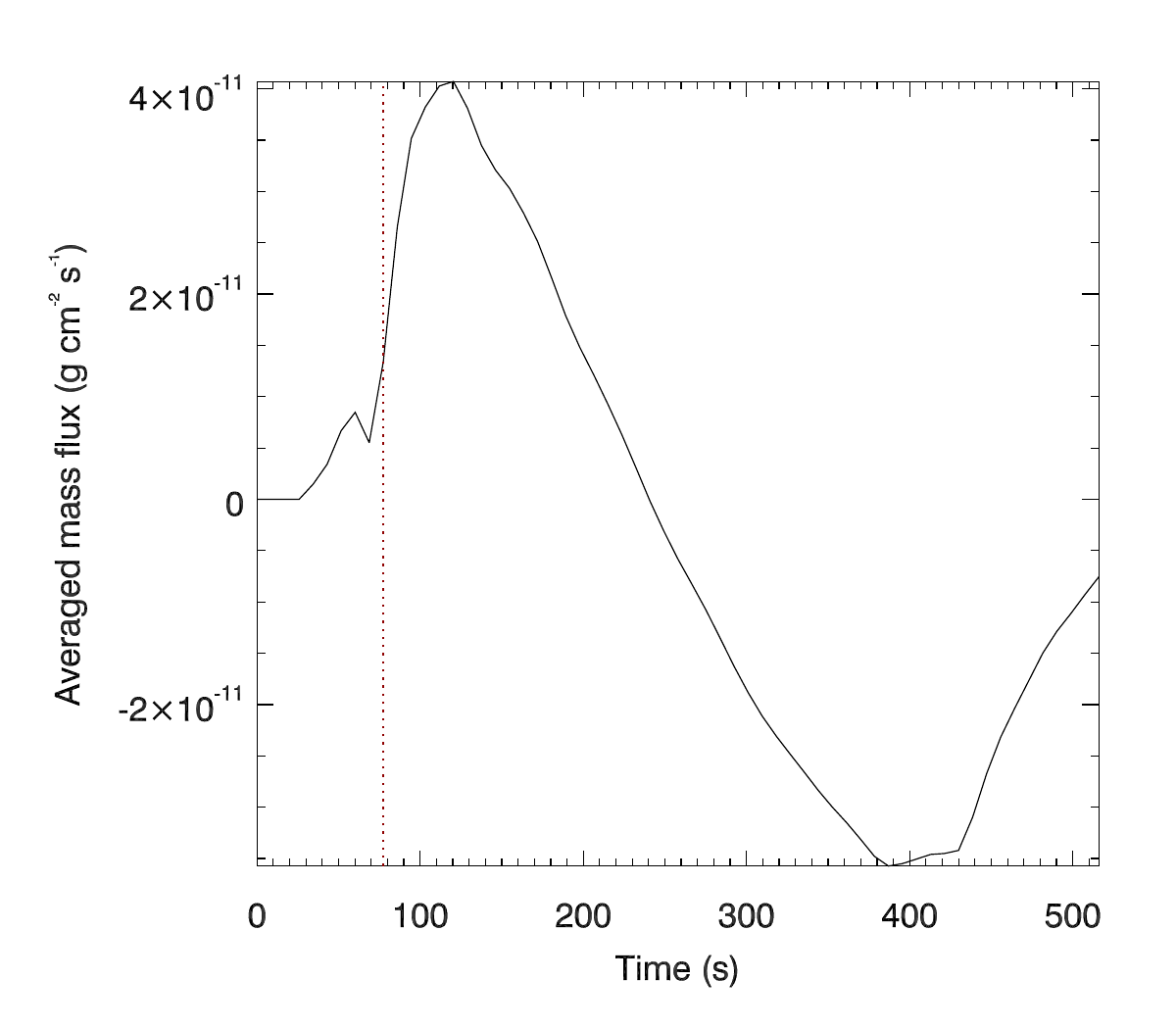}
    \caption{Averaged mass flux through the plane indicated in Figure \ref{fig3} during the whole process, the red dashed line indicates the moment reconnection occurs.} 
    \label{figa3}
\end{figure}

Furthermore, we find that the reconnection facilitates the transfer of cool materials originally included within the flux rope to the upper corona. Figure \ref{figa3} illustrates the averaged mass flux through a plane just above the reconnection region (z $\sim$ 10 Mm), with the dashed line indicating the onset time of reconnection. The mass flux is calculated by $F = \rho \boldsymbol{v}$. A significant increase in mass flux is observed after the external reconnection, and a peak is reached quickly in 40 seconds. This indicates that the effective mass transfer towards the upper corona is driven by reconnection. The effective mass transfer occurs on a timescale of approximately 5 minutes, aligning with the timescale of fan reconnection \citep{Cheng2023}. Note that our computational domain is significantly smaller in vertical extent than the coronal pressure scale height. Hence, within the framework of our model, the ejected material does not form the solar wind but falls back to the surface (after time ca. 250 s in Figure \ref{figa3}).

\end{appendix}

\end{document}